\documentclass[11pt]{article}
\usepackage{amsmath}
\usepackage{amsfonts,amssymb}

\begin{document}

\newcommand{\nl}{\nonumber\\}
\newcommand{\nnl}{\nl[6mm]}
\newcommand{\nle}{\nl[-2.5mm]\\[-2.5mm]}
\newcommand{\nlb}[1]{\nl[-2.0mm]\label{#1}\\[-2.0mm]}
\newcommand{\ab}{\allowbreak}

\newcommand{\e}{{\mathrm e}}
\newcommand{\mm}{{\mathbf m}}
\newcommand{\nn}{{\mathbf n}}
\newcommand{\zero}{{\mathbf 0}}
\newcommand{\xx}{{\mathbf x}}

\newcommand{\be}{\bes}
\newcommand{\ee}{\ees}
\newcommand{\bes}{\begin{eqnarray}}
\newcommand{\ees}{\end{eqnarray}}
\newcommand{\eens}{\nonumber\end{eqnarray}}

\renewcommand{\/}{\over}
\renewcommand{\d}{\partial}

\newcommand{\bra}[1]{\langle{#1}|}
\newcommand{\ket}[1]{|{#1}\rangle}
\newcommand{\bracket}[2]{\langle{#1}{#2}\rangle}
\newcommand{\no}[1]{:{#1}:}

\newcommand{\eps}{\epsilon}
\newcommand{\si}{\sigma}
\newcommand{\ka}{\kappa}
\newcommand{\la}{\lambda}
\newcommand{\w}{\omega}

\newcommand{\vect}{{\mathfrak{vect}}}
\newcommand{\map}{{\mathfrak{map}}}
\newcommand{\gl}{{\mathfrak{gl}}}
\newcommand{\ssl}{{\mathfrak{sl}}}

\newcommand{\tr}{{\rm tr}}
\newcommand{\oj}{{\mathfrak g}}
\newcommand{\kmg}{\widehat\oj}
\newcommand{\hatoplus}{\widehat\oplus}

\newcommand{\TT}{{\mathbb T}}
\newcommand{\RR}{{\mathbb R}}
\newcommand{\CC}{{\mathbb C}}
\newcommand{\ZZ}{{\mathbb Z}}
\newcommand{\NN}{{\mathbb N}}

\newcommand{\EE}{{\mathcal E}}

\newcommand{\xmu}{\xi^\mu}
\newcommand{\ynu}{\eta^\nu}
\newcommand{\dmu}{\d_\mu}
\newcommand{\dnu}{\d_\nu}
\newcommand{\drho}{\d_\rho}

\renewcommand{\L}{{\cal L}}
\newcommand{\J}{{\cal J}}

\newtheorem{postulate}{Postulate}
\newtheorem{theorem}{Theorem}

\title{{Three Principles for Quantum Gravity}}

\author{T. A. Larsson \\
Vanadisv\"agen 29, S-113 23 Stockholm, Sweden\\
email: thomas.larsson@hdd.se}

\maketitle
\begin{abstract}
We postulate that the fundamental principles of Quantum Gravity are 
diffeomorphism symmetry, unitarity, and locality. Local observables are 
compatible with diffeomorphism symmetry in the presence of diff anomalies,
which modify the symmetry algebra upon quantization.
We describe the generalization of the Virasoro
extension to the diffeomorphism algebra in several dimensions, 
and its off-shell representations.
These anomalies can not arise in QFT, because the 
Virasoro-like cocycles are functionals of the observer's
spacetime trajectory, which is not present in QFT.
Possible implications for physics are discussed.

\end{abstract}

\vskip 1cm

\section{The postulate}

All known physical phenomena are described by two theories: General
Relativity (GR), which describes gravity, and Quantum Field Theory
(QFT), which describes everything else. For the past 85 years,
physicists have seeked to unify these two theories into a single theory of 
Quantum Gravity (QG). Alas, GR and QFT are mutually incompatible, and
despite an immense amount of work by many leading physicists, there has
been no clear progress. In particular, the origin of mass
quantization (why is $m_p \approx 1836 \cdot m_e$?) remains a complete
mystery.

In view of this failure, I propose to take a step back and reexamine
the fundamental principles that QG should rest upon. A radical
possibility is that QG simply combines the fundamental properties of 
GR and QFT:

\begin{postulate}[Main postulate, physical version]
\label{Post:phys}
Quantum Gravity has the following properties:
\begin{enumerate}
\item
Spacetime diffeomorphism symmetry (the gravity property).
\item
Unitarity and energy bounded from below (the quantum property).
\item
Locality (the field property).
\end {enumerate}
\end{postulate}

None of the currently popular QG candidates satisfy all three
properties. There is of course an excellent reason for this: according
to standard wisdom, the three properties in the main postulate are
mutually incompatible. 

\begin{theorem}[No-go theorem, physical version]
\label{Thm:NoGoPhys}
There are no local observables in QG. In QFT, local observables
are gauge-invariant unitary operators. Since diffeomorphisms are part of the
gauge group of GR, any observable must be invariant under arbitrary
diffeomorphisms, and hence it can not be local. 
The three properties of Postulate \ref{Post:phys} are mutually
exclusive.
\end{theorem}

To gain some further insight, let us rephrase the postulate in
terms of the representation theory of the diffeomorphism group.

\begin{postulate}[Main postulate, representation theory version]
\label{Post:repr}
Quantum Gravity has the following properties:
\begin{enumerate}
\item
All objects in the theory carry representations of the spacetime 
diffeomorphism group (the gravity property).
\item
The representations are unitary and of lowest-energy type
(the quantum property).
\item
At least some representations are non-trivial (the field property).
\end{enumerate}
\end{postulate}

The no-go theorem can now be formulated as follows:

\begin{theorem}[No-go theorem, representation theory version]
\label{Thm:NoGoRep}
The spacetime diffeomorphism group has no non-trivial, proper, unitary
representations of lowest-energy type.
\end{theorem}

This theorem is correct as stated, but no theorem is stronger than 
its axioms. The keyword is ``proper''; if we relax that condition,
the theorem no longer holds, as the following example illustrates.

Consider the group of diffeomorphisms on the circle, and
its Lie algebra of vector fields $\vect(S^1) \equiv \vect(1)$; 
for brevity, the notation only indicates the number of dimensions.
The infinitesimal generators $L_m = -i\exp(imx)\d/\d x$, $m \in \ZZ$, 
satisfy 
\be
[L_m, L_n] = (n-m) L_{m+n}.
\label{vect1}
\ee
The only unitary lowest-energy representation of $\vect(1)$ is the
trivial one, in accordance with Theorem \ref{Thm:NoGoRep}. However, it is 
well known from conformal field theory (CFT) how to solve this problem.
$\vect(1)$ admits a non-trivial central extension, the Virasoro 
algebra:
\be
[L_m, L_n] = (n-m) L_{m+n} - {\frac c {12}}(m^3-m)\delta_{m+n},
\label{Vir1}
\ee
where $\delta_m$ denotes the Kronecker delta and $c$ is the central
charge. A lowest-energy representation has a unique vacuum vector 
$\ket h$, which satisfies
\bes
L_0 \ket h &=& h \ket h, 
\nle
L_{-m} \ket h &=& 0, \qquad \hbox{for all $-m < 0$}.
\eens
The Virasoro algebra has non-trivial unitary representations of 
lowest-energy type, e.g. the entire Verma modules for $c > 1, h > 0$
or the discrete unitary series \cite{DFMS94}:
\bes
c &=& 1 - {\frac 6 {m(m+1)}}, \nle
h &=& {\frac {((m+1)r-ms)^2 - 1} {4m(m+1)}}, \qquad
1 \leq r < m, 1 \leq s \leq r.
\eens
For these values of $c$ and $h$, CFT satisfies all conditions in
Postulate \ref{Post:repr}:
\begin{enumerate}
\item
The theory has a symmetry under the diffeomorphism 
group on the circle.
\item
The theory is unitary and the energy is bounded from below - the $L_0$
eigenvalue is at least $h$ for every state in the Hilbert space.
\item
The theory is local in the sense that correlation functions depend on
separation. E.g., the correlator between two primary fields behaves like
\be
\langle \phi(z) \phi(w) \rangle \approx (z-w)^{-2h}
\ee
when $z \to w$.
\end{enumerate}

It is now clear how the no-go theorem can be
avoided: allow  projective representations of the spacetime 
diffeomorphism group.

\begin{theorem}
To satisfy all desiderata in the main postulate 
it is necessary that symmetry of QG is some group 
extension of the spacetime diffeomorphism group. This converts the
classical diffeomorphism gauge symmetry into an ordinary quantum
symmetry, which does not need to commute with observables.
\end{theorem}

On the Lie algebra level, this amounts to replacing $\vect(d)$, the
Lie algebra of vector fields in $d$ dimensional spacetime, 
with a Lie algebra extension thereof. Since this extension 
generalizes the Virasoro algebra to multi-dimensional manifolds, we
call it the multi-dimensional Virasoro algebra and denote it by
$Vir(d)$; $Vir(1)$ is the ordinary Virasoro algebra.

\section{ The objections }

Replacing $\vect(d)$ with $Vir(d)$ is a drastic step, which may 
potentially lead to several objections.

\begin{enumerate}
\item
$\vect(d)$ does not possess any central extension at all when 
$d > 1$.
\item
An extension of the diffeomorphism algebra is a diff anomaly. In
QFT, there are no diff anomalies in four dimensions \cite{Bon86}.
\item
Diffeomorphisms are part of the gauge symmetries of gravity. 
In QFT observables are gauge-invariant operators, and hence all
observables must commute with diffeomorphisms.
\item
A diff anomaly is a kind of gauge anomaly, which automatically renders
the theory inconsistent.
\end{enumerate}

The first three objections are correct as formulated, but the statements
contain assumptions that are overly strong. The last objection is
manifestly false.

\begin{enumerate}

\item
The diffeomorphism algebra in $d>1$ dimensions does not possess any 
{\bf central} extension, but it does possess non-central extensions 
that reduce to the Virasoro algebra in the case $d=1$. 
$Vir(d)$ is an extension of $\vect(d)$ by its module of closed
$(d-1)$-forms. When $d=1$, a closed zero-form is a constant function
and the extension is central. When $d>1$,
the extension does not commute with diffeomorphisms, but
there are still non-trivial Lie algebra extensions.

The multi-dimensional Virasoro algebra is described explicitly in
section \ref{sec:Vir}. For a classification of abelian extensions
of $\vect(d)$ by modules of tensor fields, see \cite{Dzhu96}.

\item
There are no diff anomalies in four dimensions {\bf within the framework
of QFT}. However, the multi-dimensional Virasoro extensions described in
section \ref{sec:Vir} certainly exist. Hence there are diff anomalies 
in arbitrary dimensions, in the same sense as the Virasoro central 
charge is a conformal anomaly in two dimensions, but 
these anomalies can not arise in QFT.

The off-shell representations of $\vect(d)$ act on tensor fields and 
tensor densities. However, tensor densities are not a good starting point
for quantization when $d>1$; in higher dimensions, normal ordering
gives rise to infinities coming from unrestricted sums over spatial
degrees of freedom. Instead we must start from histories in the
space of tensor-valued $p$-jets, $p$ finite; locally, a $p$-jet is the same
as a Taylor series truncated at order $p$. Since a $p$-jet history
consists of finitely many functions of a single variable, normal
ordering can be done without introducing any infinitities.

A $p$-jet can be thought of as a regularization of the field, but not
only so. A Taylor series does not only depend on the function being
expanded, but also on the choice of expansion point, a.k.a. the 
observer's position. This is essential, because in all known 
representations of $Vir(d)$, the extension is a functional of the 
observer's trajectory. The Virasoro-like diff anomalies can not 
arise in QFT, because they depend on degrees of freedom not available.
To construct these diff anomalies, we must replace QFT with a theory
that depends on the observer's trajectory in addition to the fields.
This theory is tentatively labelled Quantum Jet Theory (QJT).

The off-shell representations of $Vir(d)$ are explicitly described
in section \ref{sec:repr}.

\item
Diffeomorphisms generate a gauge symmetry {\bf in the absense of
diff anomalies}. A gauge anomaly converts a classical
gauge symmetry into an ordinary quantum symmetry, which acts on the Hilbert
space rather than reducing it. Hence there may be local observables
in QG in the presence of diff anomalies.

\item
It is simply not true that every theory with gauge anomalies is
inconsistent. Counterexample: according to the no-ghost theorem, 
the free subcritical string can be quantized with a ghost-free spectrum
despite its conformal gauge anomaly (\cite{GSW88}, section 2.4).
A gauge anomaly simply means
that the classical and quantum theories have different symmetry groups.

This does of course not mean that every theory with a gauge anomaly 
can be rendered consistent, but the crucial consistency criterion is
unitarity, not triviality. E.g., the gauge anomalies that appear in the
standard model are related to the Mickelsson-Faddeev (MF) 
algebra\footnote{ Note that the MF algebra is substantially different from the multi-dimensional
affine algebra $Aff(d,\oj)$ described in Section \ref{sec:Aff} below.} 
\cite{Mi89}, which is
known to lack good quantum representations; more precisely, the MF
algebra has no non-trivial, unitary representations acting on a 
separable Hilbert space \cite{Pic89}. Gauge anomalies of this type must
therefore cancel, which is also the case in the standard model. In
contrast, $Vir(d)$ may well have 
non-trivial unitary representations (this is at least the case when
$d=1$), and such diff anomalies are not necessarily a sign of 
inconsistency.

Treating an anomalous gauge symmetry as a redundancy is of course
inconsistent, since it becomes an ordinary symmetry after quantization.

\end{enumerate}

\section{Multi-dimensional Virasoro algebra}
\label{sec:Vir}

Denote by $Vir(d)$ the Virasoro algebra in $d$ dimensions. In a
Fourier basis on the $d$-torus, the generators are 
$L_\mu(m)$ and $S^\mu(m)$, $m = (m_0,m_1,...m_{d-1}) \in \ZZ^d$, 
which satisfy
\bes
[L_\mu(m), L_\nu(n)] &=& n_\mu L_\nu(m+n) - m_\nu L_\mu(m+n) \nl
&+& (c_1 m_\nu n_\mu + c_2 m_\mu n_\nu) m_\rho S^\rho(m+n), 
\nl
{[}L_\mu(m), S^\nu(n)] &=& n_\mu S^\nu(m+n)
 + \delta^\nu_\mu m_\rho S^\rho(m+n),
\label{Vird}\\
{[}S^\mu(m), S^\nu(n)] &=& 0, \nl
m_\mu S^\mu(m) &=& 0.
\eens
To see that this algebra indeed reduces to the usual Virasoro algebra when
$d=1$, we notice that the condition $m_0 S^0(m_0) = 0$ implies that
$S^0(m_0)$ is proportional to the Kronecker delta, which indeed commutes
with diffeomorphisms. So the Virasoro extension is central when $d=1$
but not otherwise. Nevertheless, (\ref{Vird}) defines a well-defined and
non-trivial Lie algebra extension of $\vect(d)$ for every $d$.

The cocycle proportional to $c_1$ was discovered by Rao and Moody
\cite{RM94}, and the one proportional to $c_2$ by myself \cite{Lar91}.
We refer to $c_1$ and $c_2$ as {\it abelian charges}, in analogy with
the central charge of $Vir(1)$.

$S^\mu(m)$ can be identified with the Fourier components of a $(d-1)$-form:
\be
\Omega(m) = \eps_{\mu_1 \mu_2 ... \mu_d} S^{\mu_1}(m) dx^{\mu_2} ... dx^{\mu_d}.
\ee
The last condition in (\ref{Vird}) asserts that this $(d-1)$-form is closed.

In the sequel we will use a different formulation not specific to tori.
Let $\xi=\xmu(x)\dmu$ be a vector field, with commutator
$[\xi,\eta] \equiv \xmu\dmu\ynu\dnu - \ynu\dnu\xmu\dmu$.
The Lie derivatives $\L_\xi$ are the generators of $\vect(d)$.
$Vir(d)$ is defined by the following brackets
\bes
[\L_\xi,\L_\eta] &=& \L_{[\xi,\eta]}
 + {1\/2\pi i}\int dt\ \dot q^\rho(t)
 \Big\{ c_1\ \d_\rho\dnu\xmu(q(t))\dmu\ynu(q(t)) \ +\nl
&&\quad+\ c_2\ \d_\rho\dmu\xmu(q(t))\dnu\ynu(q(t)) \Big\}, 
\nl
{[}\L_\xi, q^\mu(t)] &=& \xmu(q(t)),
\label{DRO}\\
{[}q^\mu(t), q^\nu(t')] &=& 0.
\eens
The connection between (\ref{Vird}) and (\ref{DRO}) is given by
\bes
L_\mu(m) &=& \L_{-i\exp(im\cdot x)\dmu}, 
\nle
S^\mu(m) &=& {1\/2\pi} \int dt\ \exp(im\cdot q(t))\  \dot q^\mu(t).
\eens
In particular, the closedness condition in (\ref{Vird}) becomes 
$\int dt\ {d\/dt}(\exp(im\cdot q(t))) \equiv 0$.

\section{Off-shell representations}
\label{sec:repr}

To construct Fock representations of $Vir(1)$ is straightforward:
\begin{itemize}
\item
Start from classical fields, i.e. primary fields = scalar densities.
\item
Introduce canonical momenta.
\item
Normal order.
\end{itemize}
The first two steps of this procedure generalize nicely to higher
dimensions, but the third step leads to infinitites due to unrestricted sums
over spatial directions. This is the reason why the representations
of $Vir(d)$, $d \geq 2$, do not act on quantum fields.

Instead, we notice that $\vect(d)$ can be embedded into a Heisenberg 
algebra with $2d$ generators $q^\mu$ and $p_\nu$, and brackets
\be
[q^\mu, p_\nu] = i\delta^\mu_\nu, \qquad
[q^\mu, q^\nu] = [p_\mu, p_\nu] = 0.
\label{H1}
\ee
The embedding is given by
\be
\L_\xi = i \xmu(q) p_\mu.
\ee
Hence $\vect(d)$ acts on the corresponding Fock module, which can be
identified with the space of spacetime fields:
\be
\L_\xi \Phi(q) = \xmu(q) \d_\mu \Phi(q).
\ee

Since the Heisenberg algebra (\ref{H1}) is finite-dimensional, the 
Fock representation of $\vect(d)$ is proper. To obtain the extensions
in (\ref{Vird}), we need to find an embedding into an infinite-dimensional
Heisenberg algebra. To this end, introduce infinitely many oscillators
$q^\mu(t)$ and $p_\nu(t)$, $t \in S^1$, with non-zero brackets
\be
[q^\mu(t), p_\nu(t')] = i\delta^\mu_\nu \delta(t-t').
\label{H2}
\ee
The embedding is given by
\be
\L_\xi = i\int dt\ \xmu(q(t)) p_\mu(t),
\label{Lxi}
\ee
where the integral runs over $0 \leq t < 2\pi$.

Unlike the finite-dimensional case, the infinite-dimensional Heisenberg
algebra (\ref{H2}) has several inequivalent Fock representations.
To satisfy the quantum property, we must choose the one with energy 
bounded from below, where energy is identified with the frequency dual
to the circle variable $t$. The Fock module consists of all functions
of the positive-frequency Fourier components, plus 
half of the zero-frequency components. 

However, the operators (\ref{Lxi}) do not act in a well-defined
manner on this Fock space, because the action on the
Fock vacuum is infinite. To remove this infinity, we must normal order.
Because the oscillators $q^\mu(t)$ commute among themselves,
this amounts to moving the positive-frequency components of $p_\mu(t)$
in (\ref{Lxi}) to the left. The normal ordered-operators satsify
the multi-dimensional Virasoro algebra (\ref{DRO}) with
$c_1 = 2d$, $c_2 = 0$. $q^\mu(t)$ is the same in both (\ref{DRO}) and
(\ref{Lxi}).

More general Fock representations act on histories the the space of 
$p$-jets \cite{Lar98,Lar15}, which locally can be identified with the space of
Taylor series truncated at order $p$.
Consider a spacetime field $\phi(x)$, expand it in a Taylor
series around $q^\mu$, and truncate at order $p$.
\be
\phi(x) = \sum_{|\mm|\leq p} {1\/\mm!} \phi_\mm(x-q)^\mm,
\label{Taylor}
\ee
where $\mm = (m_0, \ab m_1, \ab ..., \ab m_{d-1})$, all $m_\mu\geq0$, is a
multi-index of length $|\mm| = \sum_{\mu=0}^{d-1} m_\mu$,
$\mm! = m_0!m_1!...m_{d-1}!$, and
\be
(x-q)^\mm = (x^0-q^0)^{m_0} (x^1-q^1)^{m_1} ... (x^{d-1}-q^{d-1})^{m_{d-1}}.
\label{power}
\ee
The space of $p$-jets is spanned by the Taylor coefficients
$\phi_\mm$, $|\mm| \leq p$ and the expansion point $q^\mu$.

Now consider $p$-jet histories by letting everything depend on an extra
circle parameter $t \in S^1$.
The Heisenberg algebra is spanned by the oscillators $q^\mu(t)$, $p_\nu(t)$, 
$\phi_\mm(t)$, and $\pi^\nn(t)$, obeying (\ref{H2}) and 
\be
[\phi_\mm(t), \pi^\nn(t')] = i \delta^\nn_\mm \delta(t-t').
\label{H3}
\ee
After normal ordering, denoted by double dots $:\ :$,
we obtain a projective Fock representation of the diffeomorphism algebra
\be
\L_\xi = i \int dt\ \Big\{ \no{\xmu(q(t)) p_\mu(t)} -
\sum_{\mm,\nn} \no{ \pi^\nn(t) T^\mm_\nn(\xi(q(t))) \phi_\mm(t) } \Big\},
\label{Lxino}
\ee
where the sum runs over all $\mm$ and $\nn$ such that
$|\mm|\leq|\nn|\leq p$. $T^\mm_\nn(\xi)$ are some functions of $\xi^\mu$ and its 
derivatives up to order $p+1$, explicitly written down in \cite{Lar98}.

The construction is readily generalized to fermionic fields, but the
expansion point $q^\mu$ is of course always bosonic.

A major shortcoming of this construction is that only linear representations
have been considered. In physics, we are ultimately interested in
unitary representations, but we have nothing to say about that.

The jet data have a natural physical interpretation. The expansion
point $q^\mu$ is the observer's position and $p_\mu$ his
momentum. The Taylor coefficients $\phi_\mm$ are the excitations of the field
that the observer can measure with a local detector.

\section{ Reparametrizations and the messy cocycles}
\label{sec:mess}

By passing to histories in the space of $p$-jets, the Taylor series 
(\ref{Taylor}) is replaced by a field that depends both on the spacetime
coordinates $x^\mu$ and the trajectory parameter $t$:
\be
\phi(x,t) = \sum_{|\mm|\leq p} {1\/\mm!} \phi_\mm(t) (x-q(t))^\mm.
\label{Taylor2}
\ee
The natural algebra that acts on such a field is $\vect(d)\oplus\vect(1)$, 
where the first factor describes spacetime diffeomorphisms and the second 
reparametrizations of the observer's trajectory. We can thus enlarge the
representations in the previous section to this larger algebra, by 
embedding the extra $\vect(1)$ generators into the same Heisenberg algebra.

Upon normal ordering, $\vect(d)\oplus\vect(1)$ acquires four different
cocycles and becomes $Vir(d)\hatoplus Vir(1)$. $Vir(d)$ has two cocycles,
$Vir(1)$ has one, and the final cocycle is found in the cross term, which
is indicated by the notation $\hatoplus$. Explicit formulas can be found
in \cite{Lar98}.

Whereas this is the mathematically clean formulation, it is somewhat redundant
from a physical point of view. There are two different time coordinates: 
the $x^0$ coordinate and the parameter $t$. To bring out the physical
content, we can use a trick described in \cite{Lar98}. 
Before normal ordering, we may pretend that all brackets are Poisson brackets.
To eliminate reparametrizations, we impose the constraint that
all $\vect(1)$ generators vanish:
\be
L(t) \approx 0.
\ee
Supplement this first-class constraint with a gauge condition to make it
second class:
\be
q^0(t) \approx t.
\label{gaugecond}
\ee
Finally, the constraint is eliminated by replacing Poisson bracket by
Dirac brackets.
It turns out that the diffeomorphism generators $\L_\xi$ still satisfy
an extension of $\vect(d)$\footnote{ That the regular terms still 
generate $\vect(d)$ is non-trivial and can be seen by an explicit 
calculation. } by four different cocycles. 

The number of cocycles must be exactly four, because this is how many 
cocycles $\vect(d)\oplus\vect(1)$ has. Apart from the two cocycles 
in (\ref{DRO}), there are two very complicated cocycles, which are
anisotropic in the sense that the dependence on the 
zeroth coordinate is different from the others. 
They are colloquially known as the {\em messy cocycles}.
The anisotropy clearly comes from the gauge condition (\ref{gaugecond}). 
For an explicit description of the messy cocycles, see \cite{Lar97}.

If we now also set $x^0 = t$, the field (\ref{Taylor2}) becomes independent
of all Taylor coefficients $\phi_\mm(t)$ with $m_0 > 0$, because that
term is proportional to 
\be
(x^0 - q^0(t))^{m_0} = (t-t)^{m_0} = 0.
\ee
We thus have eliminated the $x^0$ coordinate altogether, and constructed
a field $\phi(\xx,t)$ which depends on a single time coordinate $t$ and
the spatial coordinates $\xx$. The price is that the projective action of 
$\vect(d)$ on the Fock space is very complicated.

\section{ Multi-dimensional affine algebra }
\label{sec:Aff}

There is an analogous multi-dimensional affine algebra
(\cite{PS88} section 4).
Let $\map(d,\oj)$ be the algebra of maps from $d$-dimensional space to
a Lie algebra $\oj$ with basis $J^a$, structure constants $f^{abc}$, 
and Killing metric $\delta^{ab}$. 
$Aff(d,\oj)$ is defined by the brackets
\be
{[}\J_X, \J_Y] &=& \J_{[X,Y]} - {k\/2\pi i}\ \delta^{ab}
 \int dt\ \dot q^\rho(t)\drho X_a(q(t))Y_b(q(t)), 
\label{Affd}
\ee
where $X = X_a(x) J^a$ is a $\oj$-valued function.
In the Fourier basis, this becomes
\be
[J^a(m), J^b(n)] = if^{abc} J^c(m+n) - k \delta^{ab} m_\mu S^\mu(m+n).
\ee
The $Aff(d,\oj)$ generators commute with $q^\mu(t)$ and $S^\mu(m)$,
and admit an intertwining action of $Vir(d)$.

Note that this cocycle is proportional to the second Casimir operator.
$Aff(d,\oj)$ is thus unrelated to the gauge anomalies appearing in the
standard model, which are proportional to the third Casimir.

Off-shell representations of $Aff(d,\oj)$ are constructed in analogy 
with $Vir(d)$. Let $M^a$ denote matrices in some 
finite-dimensional representation of $\oj$. The following expression
defines an embedding of $Aff(d,\oj)$ into the Heisenberg algebra:
\be
\J_X = 
-i\int dt\ \sum_{\mm,\nn} \no{ \pi^\nn(t) J^\mm_\nn(X(q(t))) \phi_\mm(t) }, 
\label{JX}
\ee
where
\be
J^\mm_\nn(X) &=& {\nn\choose\mm} \d_{\nn-\mm} X_a M^a.
\ee
Hence $Aff(d,\oj)$ acts on the Fock space.

Whereas the off-shell representations of $Vir(d)$ are only understood
at the linear level, unitary representations of $Aff(d,\oj)$ are easily 
constructed. Specialize (\ref{JX}) to zero-jets:
\be
\J_X &=& \int dt\ X_a(q(t)) J^a(t),
\label{JX2}
\ee
where
\be
J^a(t) = -i \no{ \pi^\zero(t) M^a \phi_\zero(t) }.
\ee
To verify that (\ref{JX2}) satisfies $Aff(d,\oj)$, it suffices to prove
that the operators $J^a(t)$ satisfy the affine algebra $Aff(1,\oj) = \kmg$.
Conversely, we obtain an $Aff(d,\oj)$ representation for every
$\kmg$ representation. If this representation is unitary, so is the 
$Aff(d,\oj)$ representation (\ref{JX2}), since $\J_X$ is merely a linear
combination of unitary operators. 

Note that we could replace $q^\mu(t)$ with a c-number because 
$[\J_X, q^\mu(t)] = 0$. This is not possible when diffeomorphisms are taken
into account, so a similar trick is not possible for $Vir(d)$.

The representation (\ref{JX2}) should be equivalent to a representation
induced from $\kmg$, living on the loop $q^\mu(t)$.
As such, it is presumably covered by the discussion in \cite{PS88}, Section 9.1.
Pressley and Segal found this result rather disappointing, 
but the reason why the irreps only require zero-jets is that the current
algebra does not explore neighboring spacetime points. The
circle $q^\mu(t)$ commutes with everything in sight and can therefore be
replaced with a c-number; the extension becomes central.
In physics, there is always an intertwining action of diffeomorphisms or
some subgroup thereof, such as the Poincar\'e group. Once such spacetime
groups are taken into consideration, $q^\mu(t)$ becomes an operator,
the extension is no longer central,
and interesting representations depend on more than $\kmg$.

\section{ Conclusion }

Locality is compatible with diffeomorphism symmetry, but only in the
presence of diff anomalies. It is necessary to quantize histories in 
the space of $p$-jets, rather than quantizing the fields themselves.
The appropriate name for this quantum theory is Quantum Jet Theory (QJT). 

Some attempts to formulate the physical consequences of QJT have been made
\cite{Lar04,Lar06}, but the results were inconclusive. Nevertheless, we can
make some general observations based on the structure of the off-shell 
representations.

\begin{itemize}
\item
Since diff anomalies of the type in (\ref{Vird})
can not arise in QFT, QJT is substantially different from QFT.

\item
Passage to $p$-jets is a kind regularization, because a problem in QFT 
is replaced by a problem with fewer degress of freedom.

\item
We ultimately want to remove the regulator, which amounts to taking the
jet order $p\to\infty$. The abelian charges diverge in this limit.
It is not surprising that the field theory problems resurface in the field
theory limit.

\item
The abelian charges are polynomials in $p$ of order $d$, where $d$ is the
dimension of spacetime. The leading terms can be cancelled in representations
acting on several fields, both bosonic and fermionic, making the total
abelian charges finite when $p\to\infty$ \cite{Lar01}.

\item
There are intriguing hints that cancellation of the leading terms in the 
abelian charges works best in four dimensions \cite{Lar04,Lar06}. 

\item
QJT is the unique regularization that preserves diffeomorphism symmetry
exactly. 

\item
The anomalies arise because QJT does not preserve the dynamics. If the 
equations of motion have order $n$, only Taylor coefficients of order
up to $p-n$ have dynamics. The rest have equations of motion which involve
Taylor coefficients of order higher than $p$, and thus lie outside $p$-jet
space.

\item
QJT is more than a regularization of QFT, because a Taylor series depends 
not only on the field being expanded, but also on the expansion point
$q^\mu$. This is naturally identified as the observer's position in spacetime.

\item
There are two types of observables in QJT: the Taylor coefficients, which are
field excitations confined to a local neighborhood of the observer, 
and the observer's position. Both are operators, measured 
by some detectors and subject to quantum fluctuations.
To obtain the field at a fixed point, both the field 
inside the detector and the detector's location are needed.
The second measurement introduces an extra fuzziness not present in QFT.

\item
The uncertainty in the observer's position is eliminated if the observer's
inert mass $M$ diverges, since the bracket between the observer's position 
and velocity is proportional to $\hbar/M$ (non-relativistically). 
In this limit QJT should be physically equivalent to QFT.

\item
Making the observer's inert mass infinite leads to problems with gravity,
since inert mass equals heavy mass. In my opinion, this is the physical
reason why QFT and GR are incompatible; they tacitly make incompatible 
assumptions about the observers mass:
\begin{itemize}
\item
In GR, the observer's heavy mass is assumed to vanish, so
the observer does not disturb the fields.
\item
In QFT, the observer's inert mass is assumed to be infinite, so the fields
do not disturb the observer. 
\end{itemize}

\item
The abelian charges are generalizations
of the Virasoro central charge, which is known to couple to length
scales. E.g. in Einstein�s gravity in three-dimensional AdS space, 
the central charge $c = -3\ell/2G$, where $\ell$ is the AdS radius 
and $G$ is Newton�s constant \cite{BH86}. By analogy, we expect 
that the abelian charges are manifested in the large-scale
structure of the universe, e.g. as a cosmological constant.

\item
If QG has local observables, any result that assumes that QG is a non-local
theory is questionable.

\end{itemize}

\end{document}